\documentclass[twocolumn,pre]{revtex4}

\usepackage{eurosym}
\usepackage{graphicx}
\usepackage{epstopdf}
\usepackage{amsmath}
\usepackage{latexsym}
\usepackage{epsfig}
\usepackage{makeidx}
\usepackage{amsfonts}

\usepackage{bm}
\newcommand{\vect}[1]{\boldsymbol{\mathbf{#1}}}

\begin{document}

\title{Memory effects and L\'evy walk dynamics in intracellular transport of cargoes}
\author{Sergei Fedotov$^1$, Nickolay Korabel$^1$, Thomas A. Waigh$^{2,3}$, Daniel Han$^{1,4}$ and Victoria J. Allan$^4$}
\affiliation{$^1$ School of Mathematics, University of Manchester, Manchester M13 9PL, UK}
\affiliation{$^2$ Biological Physics, School of Physics and Astronomy, University of Manchester, Manchester M13 9PL, UK}
\affiliation{$^3$ The Photon Science Institute, University of Manchester, Manchester M13 9PL, UK}
\affiliation{$^4$ Faculty of Biology, Medicine and Health, School of Biological Sciences, University of Manchester, Manchester M13 9PL, UK}

\begin{abstract}

We demonstrate the phenomenon of cumulative inertia in intracellular transport
involving multiple motor proteins in human epithelial cells by measuring the empirical
survival probability of cargoes on microtubules and their detachment rates.
We found the longer a cargo moves along a microtubule, the less likely it detaches from it.
As a result, the movement of cargoes is non-Markovian and involves a memory.
We observe memory effects on the scale of up to 2 seconds. We provide a theoretical
link between the measured detachment rate and the super-diffusive L\'evy
walk-like cargo movement.

\end{abstract}

\maketitle



\begin{figure}[t]
\centerline{ \psfig{figure=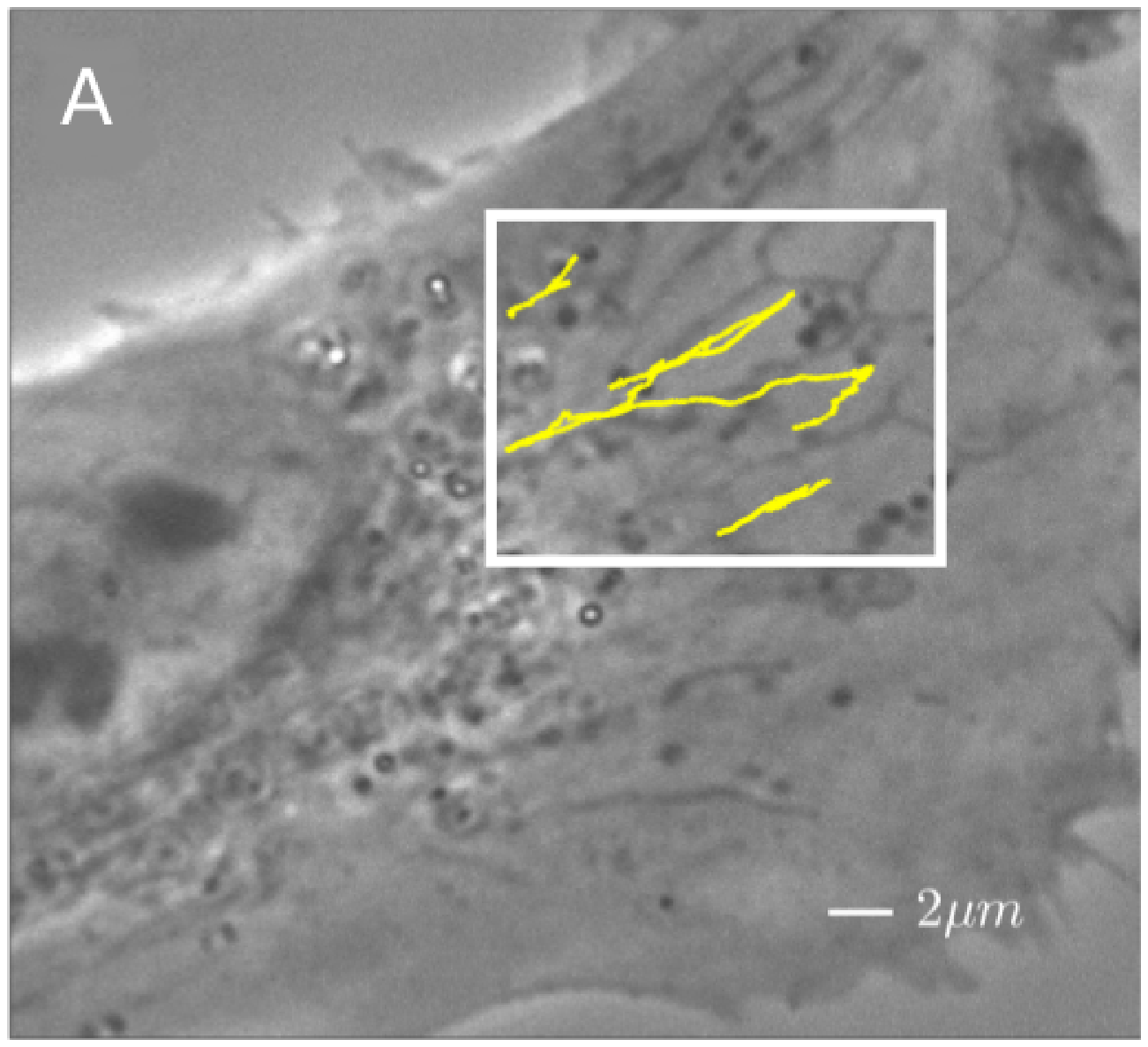,width=60mm,height=55mm} }
\centerline{ \psfig{figure=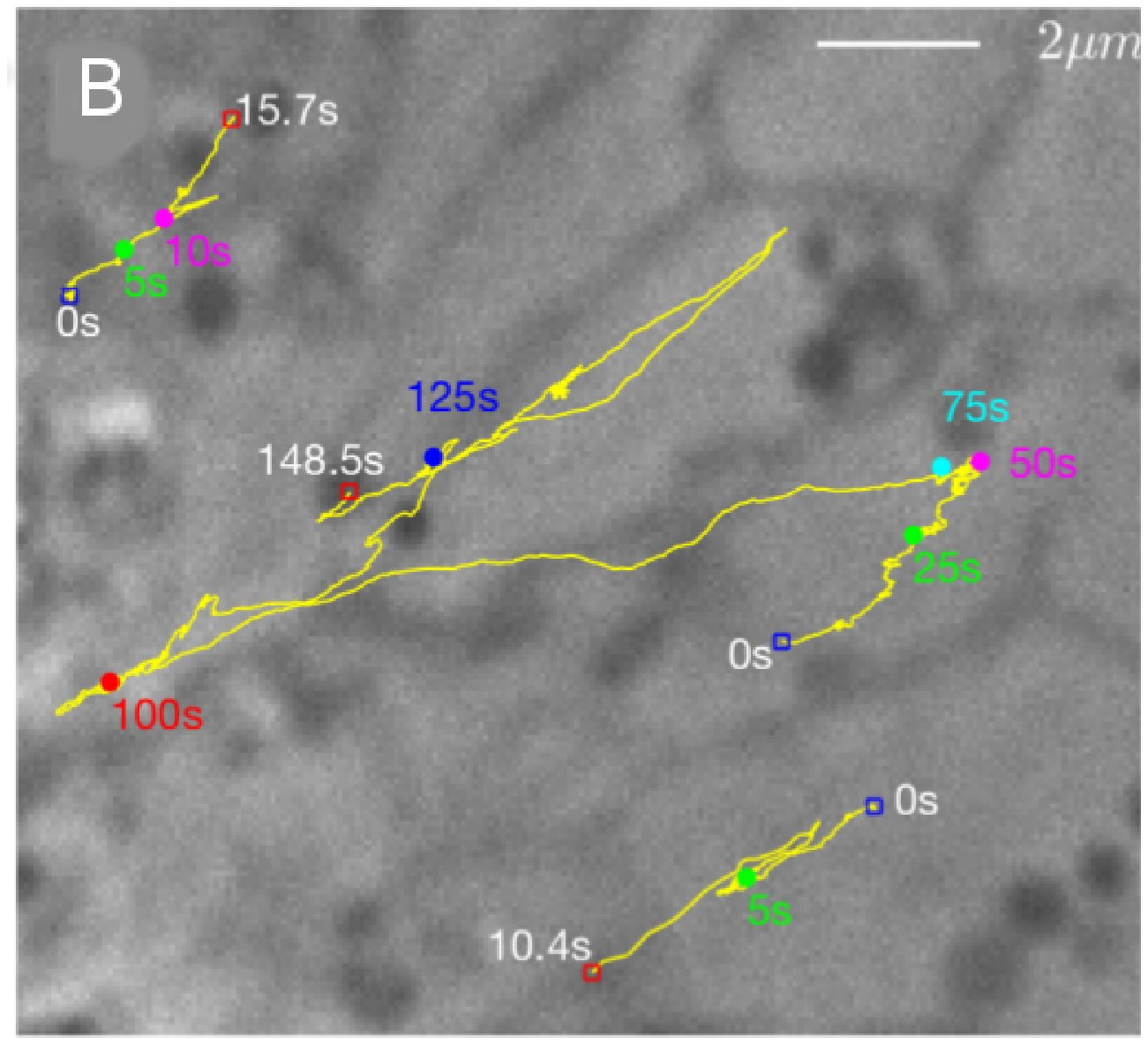,width=60mm,height=55mm} }
\caption{Cargoes movement along microtubules visualized in a living RPE cell (panel A). Panel B
shows an enlargement of the boxed region in A. Trajectories (yellow) consists of long flights in one direction and multiple turnings.
Numbers indicate the time progress. The longest trajectory ($148.5$ s) in the panel
B consists of $900$ flights (see Sec.\ V for the description of the segmentation procedure).}
\label{FIG1}
\end{figure}

\section{Introduction} \label{Sec1}

Intracellular transport of cargoes along microtubules is a classical example of active transport \cite{Valle, Bressloff}.
It is critical to cellular function and it is a challenging statistical problem from the viewpoint of active matter physics 
\cite{Loverdo, Wang, Trong, Brangwynne, Tabei}. 
{\it In-vitro} experiments show that the distance travelled by cargoes substantially increases when the cargoes
are transported by multiple motor proteins \cite{Vershinin}. Various models have been developed that aim to explain how motors
achieve long range transport along microtubules \cite{Lipowsky, Berger11, Berger12, 
Bressloff1, Julicher, Kolomeisky, WelteGross98, MullerKlumppLipowsky, Leidel, Hancock, Harrison, Ajay}.

Recently, it has been discovered that active cargo transport
{\it in-vivo} self-organizes into L\'evy walks \cite{Granick}. 
L\'evy walks describe a wide spectrum of biological processes, such as T-cells migrating in 
brain tissue, collective behaviour of swarming bacteria and animals optimizing their search for sparse food \cite{Zaburdaev}.
Endosomal L\'evy dynamics involves long flights in one direction due to the active movement along microtubules
driven by multiple motors. When all active motors disengage, the cargo complexes detach from the
microtubule and reattach to a new microtubule heading in another direction \cite{Granick}.
The travel distances have power-law distributions with diverging variances
\cite{MetzlerKlafter, Barkai, HF, Sokolov, Zaburdaev}
that determine the anomalously long flights of cargo complexes.
In Ref.\ \cite{Granick} the authors proposed the concept of memoryless self-reinforced directionality
to demonstrate the emergence of L\'evy walks.\

However, the unanswered question remains: what is the precise
mesoscopic kinetic mechanism of anomalous directional persistence?
To answer this question, we performed {\it in vivo} experiments recording thousands of trajectories
of intracellular lipid bound vesicles in live retinal pigment epithelium (RPE) cells and
human bone osteosarcoma epithelial (U2OS) cells. We found similar results for both cell lines. Therefore,
we report only results for RPE cells since a microscope with higher resolution was used to image them
(see Section \ref{Sec5} for experimental details). In Fig.\ \ref{FIG1}, we illustrate L\'evy-like trajectories of vesicles inside RPE cells which consist of
long persistent runs in one direction separated by rapid jiggling events when vesicles change direction.\

In this paper we reveal a new mechanism for anomalous directional 
persistence of cargoes in human cells:  {\it the phenomenon of cumulative inertia}.
Experimentally we found the longer a cargo moves along a microtubule, the less likely it will detach from it. 
To our knowledge our data provides the first direct measurement of the mesoscopic detachment rate
as a decreasing function of the running time. We found that this time follows a heavy tailed Pareto distribution
which leads to a L\'evy walk-like movement of cargoes. Since the observed detachment probability depends on how long the cargo
has been moving, this active transport involves memory and it exhibits a typical non-Markovian behaviour.
Note that we are dealing with the memory which is not physically or chemically 
stored or retrieved and therefore energetically costs the cells nothing.

\section{Empirical survival probability, mesoscopic detachment rate and mean residual time} \label{Sec2}

One of our aims is to measure important
statistical characteristics of cargo transport: the empirical survival probability of cargoes on the microtubule,    
the empirical mesoscopic detachment rate and the mean residual time. 

\begin{figure}[t]
\centerline{ \psfig{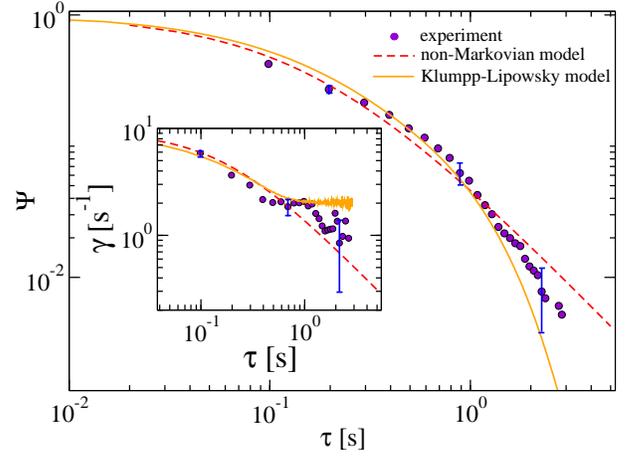} }
\caption{The experimentally determined survival function $\Psi(\tau)$ (dots) as a function of time $\tau$ 
decays as a power law on intermediate time scales with an anomalous exponent $\alpha=1.6 \pm 0.17$ 
and a time scale parameter $\tau_0=0.17 \pm 0.1$ s.\ 
The errors were calculated with the Greenwood formula for the non-parametric Kaplan-Meier estimator \cite{Aalen}.
The red dashed curve is a power law fit 
Eq.\ (\ref{Psi}) with the same $\alpha$ and $\tau_0$. The orange solid curve is the survival 
function obtained in numerical simulations of the Klumpp-Lipowsky model of the cargo dynamics with $6$ kinesin and $6$ 
dynein motors. Details of simulations are given in the section \ref{Sec6}. 
Inset: the corresponding empirical mesoscopic detachment rate function $\gamma(\tau)$ (dots) 
as a function of time is inversely proportional to $\tau$, Eq.\ (\ref{gamma}), 
on intermediate time scales (red dashed curve) with the same $\alpha$ and $\tau_0$ 
as in the main figure. The orange solid curve is the rate function obtained in numerical simulations.}
\label{FIG2}
\end{figure}

The empirical survival probability defines the probability that the cargo have not detached from the microtubule (survived on the microtubule) beyond a specified time \cite{Aalen}. It is a common tool in many areas such as Engineering and Medicine 
to measure the time-to-failure of machine parts or the fraction of patients living for a certain amount of time after treatment \cite{Aalen}. 
The survival probability was estimated by using the non-parametric Kaplan-Meier estimator \cite{Aalen}. 
We found a good agreement between the empirical survival probability $\Psi$ and the heavy tailed Pareto distribution:
\begin{equation}
\Psi(\tau)= (1+\tau/\tau_0)^{-\alpha},
\label{Psi}
\end{equation}
with the anomalous exponent $\alpha=1.6 \pm 0.17$ and $\tau_0=0.17 \pm 0.1$ s up to $1-2$ seconds (Fig.\ \ref{FIG2}).
The empirical mesoscopic detachment rate is also found to be a decreasing function of the flight time $\tau$ (inset Fig.\ \ref{FIG2}).
This means that the longer a cargo remains on a microtubule, the less likely it will detach from it
({\it the phenomenon of cumulative inertia}). Surprisingly, although having very different origin, 
similar effect of cumulative inertia is well studied and got revived interest 
in social and behavioural sciences \cite{Morrison,FS}. 
The empirical mesoscopic detachment rate has a good fit with the
rate inversely proportional to the flight time $\tau$:
\begin{equation}
\gamma(\tau) = \alpha/(\tau + \tau_0),
\label{gamma}
\end{equation}
with the same anomalous exponent $\alpha=1.6 \pm 0.17$ and $\tau_0=0.17 \pm 0.1$ s.
Note the relationship between $\Psi(\tau)$ and  $\gamma(\tau)$:  
$\Psi(\tau)=\exp\left(-\int_{0}^{\tau} \gamma(t) dt \right)$.
The time dependent detachment rate $\gamma(\tau)$ has the following meaning:
the product $\gamma(\tau) \Delta \tau$ defines the conditional probability of
cargo detachment in the interval $(\tau, \tau+\Delta \tau)$ given that it has moved along the
microtubule in the time interval $(0,\tau)$. For memoryless cargoes this rate will be constant and
will not depend on how long the cargo has moved before.
It is well-known that the empirical rate is notoriously difficult to estimate \cite{Aalen}, since it contains
the derivative of the empirical survival function. Therefore, the empirical survival probability
(which is an integral quantity) has a smoother behaviour compared to the empirical mesoscopic detachment rate function (Fig.\ \ref{FIG2}).

\begin{figure}[t]
\centerline{ \psfig{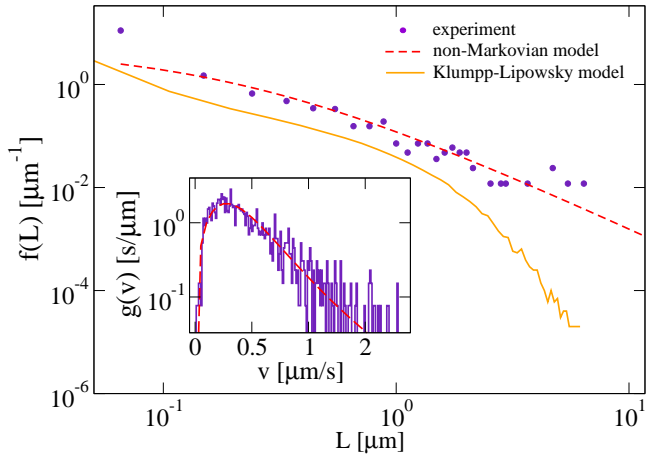} }
\caption{Main panel: The experimental flight length probability density $f(L)$ (blue dots)
fitted with Eq.\ (\ref{fL}) (red dashed curve) with $\alpha=1.5 \pm 0.17$, $\tau_0=0.33 \pm 0.1$ s and 
the average velocity of the cargo $v=0.8$ $\mu$m/s.
The flight length distribution obtained in numerical simulations of the cargo dynamics using the Klumpp-Lipowsky model
with $6$ dynein and $6$ kinesin motors (orange solid curve) decays exponentially for $L>1$ $\mu$m.
Parameters of the simulations are given in the section \ref{Sec6}. The inset shows the distribution of experimental flight
velocities $g(v)$ (blue solid curve) approximated with the Burr density (red dashed curve)
with parameters $\tilde{\gamma}=0.57\pm0.07$, $\tilde{c}=2.09\pm0.09$, $\tilde{k}=2.00\pm0.34$.
The average velocity of the cargo is $v=0.8$ $\mu$m/s.}
\label{FIG3}
\end{figure}

The value of experimental exponent $\alpha=1.6 \pm 0.17$ falls in the interval $1<\alpha<2$.
This is an extraordinary finding since it shows that the survival function has a finite mean,
$\left<T\right>=\int_0^{\infty} \tau \Psi(\tau) d\tau$, but
divergent second moment \cite{Zaburdaev}. Specifically the lack of the second moment
leads to the emergence of the L\'evy walk-like trajectories of vesicles (Fig.\ \ref{FIG1}).
Such trajectories exhibit sub-ballistic super-diffusive behaviour. We also obtain a good power
law fit with the anomalous exponent ($\alpha\simeq1.5\pm 0.17$) for the probability density of flight lengths
(Fig.\ \ref{FIG3}) which confirms the L\'evy walk nature of the vesicles motion.
The velocity $v$ of each flight was assumed to be constant. 
The probability density of flight length is obtained from the
survival function as: 
\begin{equation}
f(L) = - \Psi^{\prime}\left(L/v\right)/v,
\label{fL}
\end{equation}
where $\Psi'(z)=d\Psi(z)/dz$. In Fig.\ \ref{FIG3} the distribution of flight velocities is also shown.

\begin{figure}[t]
\centerline{ \psfig{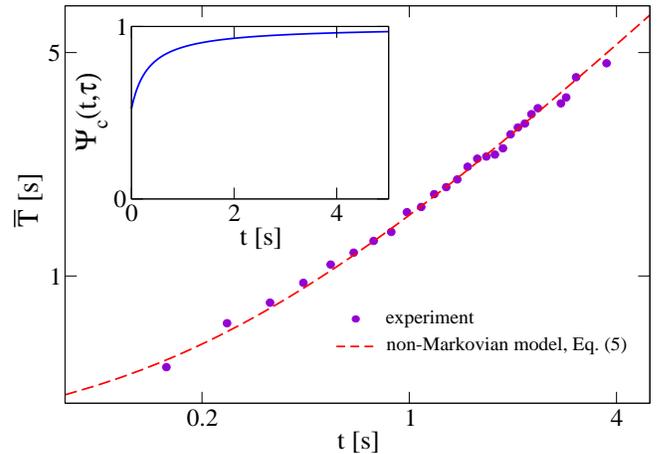} }
\caption{Main panel: The experimental mean residual time $\bar{T}$ of flights 
(blue dots) linearly increases (notice logarithmic scales on both the horizontal and vertical axes)
with the time $t$ already travelled. The theoretical prediction Eq.\ (\ref{Tt}) with parameters
$\tau_0=0.24 \pm 0.1$ s and $\alpha=1.8 \pm 0.17$ (red dashed curve) is in good agreement
with the experiment. Inset: Illustration of the increasing conditional survival function
$\Psi_{c}(t,\tau)$ given by Eq.\ (\ref{Psit}) with $\tau_0=0.2$ s, $\alpha=1.8$ and $\tau=0.1$ s. }
\label{FIG4}
\end{figure}

If the cargo has survived on the microtubule up to time $t$, how much longer is it expected to move (survive) along the microtubule? 
This time is called the mean residual time $\overline T(t)$ - another important quantitative measure of cumulative inertia. 
{\it In vivo} experiments we found that $\overline T(t)$ increases linearly in time $t$
already travelled, see Fig.\ \ref{FIG4}. The longer the cargo remains on the microtubule, the larger the
mean residual time, so the inertia is accumulated. This behaviour is drastically different to
memoryless systems where $\overline T(t)$ is constant and does not depend on the prehistory.
The data can be well explained by the conditional survival function for a random attachment time $T$,
$\Psi_{c}(t,\tau)=\Pr \left\{ T> t+\tau |T>t \right\}.$ In our case:
\begin{equation}
\Psi_{c}(t,\tau)=\frac{\Psi(t+\tau)}{\Psi(t)}= \left(\frac{\tau_{0}+t}{\tau _{0}+t+\tau}\right) ^{\alpha},
\label{Psit}
\end{equation}
is an increasing function of time $t$ already spent on the microtubule for a fixed $\tau$.
The behaviour of $\Psi_{c}(t,\tau)$ is illustrated in the inset of Fig.\ \ref{FIG4}.
The mean residual time $\overline T(t)$ can be obtain as \cite{Morrison78}:
\begin{equation}
\overline T(t)=\int_{0}^{\infty }\Psi_{c}(t,\tau)d\tau =\frac{\tau _{0}+t}{\alpha -1}.
\label{Tt}
\end{equation}
We found a good agreement between experimental mean residual time $\overline T(t)$ and 
Eq.\ (\ref{Tt}) with $\tau_0 = 0.24 \pm 0.1$ and the anomalous exponent $\alpha =1.8 \pm 0.17$, 
Fig.\ \ref{FIG4}. Notice that the values of anomalous exponents obtained by fitting experimental data $\alpha=1.6\pm0.17$ in Fig. \ref{FIG2}, 
$\alpha=1.5\pm0.17$ in Fig. \ref{FIG3} and $\alpha=1.8\pm0.17$ in Fig. \ref{FIG4} agree with each other within the error bars. 
This cumulative inertia with
$1<\alpha<2$ explains the dramatic increases of the travelled distance that are typical for L\'evy walk.
Since this is a non-Markovian effect with memory, our explanation of anomalous long distance
transport is completely different from the idea of memoryless self-reinforced directionality \cite{Granick}
when the probability $P(L)$ of traveling in some direction grows with the distance $L$ already travelled.
This probability can be obtained in terms of the conditional survival function $\Psi _{c}(t,\tau)$
as $P(L)=\Psi_{c}(L/v,\tau)$.

\section{Microscopic mechanism for the emergence of the decreasing mesoscopic detachment rate} \label{Sec3}

The question arises: what is the microscopic mechanism of the decreasing mesoscopic detachment rate $\gamma(\tau)$?
A first insight could be obtained from a classical microscopic Klumpp-Lipowsky model \cite{Lipowsky}. 
Note that the authors of Ref.\ \cite{Lipowsky} did not consider this time dependent rate. 
Instead, they found a constant effective unbinding rate. 

Consider a cargo which is pulled by multiple motors.
We assume that initially the cargo attaches to the microtubule with a single
motor. As the cargo moves along the microtubule, the number of engaged motors 
$N(t)$ varies from $1$ to $\bar{N}$.  
Motors attach to the microtubule and detach from it with effective microscopic rates $\pi_n$ and $\epsilon_n$ ($n$ is the number of engaged motors). 
We define the random detachment time $T$ as 
the time when all active motors together with the cargo detach from
the microtubule. In other words we have a random walk in the space of the number of engaged motors. 
In order to obtain the effective detachment rate $\gamma(\tau)$,
one can define the survival function $\Psi(\tau)$ of the cargo to remain on the microtubule as the
probability $\Psi(\tau)=\text{Pr}\{T>\tau\}=1-\text{Pr}\{N(\tau)=0 | N(0)=1 \}$ \cite{Lipowsky}.
The effective detachment rate is defined as \cite{Cox}:
\begin{equation}
\gamma(\tau)=-\Psi'(\tau)/\Psi(\tau).
\label{gammaDef}
\end{equation}
The function $\Psi (\tau )$ can be written in terms of 
the probability distribution of the first passage time $
T$ from the cargo state with $1$ engaged motor at time $0$ to the
cargo state with zero engaged motors:
\begin{equation}
F(\tau)=\Pr \left\{ T<\tau \right\} 
\end{equation}%
as
\begin{equation}
\Psi (\tau )=1-F(\tau).
\label{sur}
\end{equation}%
To find the distribution function  $F(\tau)$ one can introduce the transition probability \cite{Lipowsky}
\begin{equation}
P_{n,0}(\tau)=\Pr \left\{ N(\tau)=0|N(0)=n\right\} .
\end{equation}%
Define $F(\tau)$ as
\begin{equation}
F(\tau)=P_{1,0}(\tau),
\end{equation}%
where $P_{1,0}(\tau )$ is the transition probability from the cargo state
with $1$ engaged motor to the state with zero engaged motors. $P_{n,0}(\tau)$ obey the
system of the backward Kolmogorov equations \cite{Cox}:
\begin{equation}
\frac{dP_{n,0}(t)}{dt}=-\left( \varepsilon _{n}+\pi _{n}\right)
P_{n,0}(t)+\varepsilon _{n}P_{n-1,0}(t)+\pi _{n}P_{n+1,0}(t)
\label{back}
\end{equation}%
with the initial conditions $P_{n,0}(0)=0$ for $2\leq n\leq \bar{N},$ $P_{0,0}(t)=1
$ ($n=0$ is the absorbing state). Here $\varepsilon _{n}$ and $\pi _{n}$ are
the unbinding and binding rates. One can solve these equations and find the transition probability
$P_{1,0}(\tau )$. It follows from Eqs. (\ref{gammaDef}) and (\ref{sur}) that 
the effective detachment rate is
\begin{equation}
\gamma(\tau) = \frac{P^{\prime}_{1,0}(\tau)}{1-P_{1,0}(\tau)}.
\label{rate}
\end{equation}%
This mesoscopic detachment rate is essentially different from
the effective constant unbinding rate obtained within the Klumpp-Lipowsky model in Ref.\ \cite{Lipowsky} from equilibrium conditions. 
It is time-dependent and decreasing function of $\tau$. 
Our purpose is to show that
the rate $\gamma(\tau )$ is a decreasing function of the running time $\tau $. It means that the
longer a cargo moves along the microtubule, the smaller is the probability
that it will detach and switch the direction in the next time interval (a cumulative inertia effect). 
Numerical modelling confirms that $\gamma$ is a decreasing function of $\tau$ on a certain time scale
(see the orange solid curve in the inset of Fig. 2).

{\it In vitro} experimental data indicates that adding just one extra motor increases
cargo run lengths by at least one order of magnitude compared to the distance traveled by a single motor \cite{Vershinin}. 
To understand intuitively the reason why $\gamma(\tau)$ decreases with the flight time $\tau$,
consider the first event of attachment of cargo complex with one motor. Initially the rate $\gamma(0)=\epsilon_1$. 
Note that this defines the microscopic time scale $1/\epsilon_1$. 
In turn, the mesoscopic time scale is given by 
$\left<T\right>=\int_0^{\infty} \tau \Psi(\tau) d\tau$, $\left<T\right> \gg 1/\epsilon_1$. 
If the second motor attaches before the first motor detaches, the load is shared 
between two motors and the detachment rate $\epsilon_2$ decreases, $\epsilon_2<\epsilon_1$.
As a result of this stochastic dynamics, the number of participating motors increases and therefore
the detachment probability of the cargo decreases with the flight time. Cumulative inertia occurs due to
multiple attachment and reattachment of motors before the cargo finally detaches from the microtubule.
This leads to a dramatic increase of the travelled distance due to the directional persistence \cite{Vershinin}.

We consider the cargo pulled by two motors $\bar{N}=2$ and show how the essential improvement of travel distance occurs. 
The backward Kolmogorov equations (\ref{back}) take the form:
\begin{eqnarray}
\frac{dP_{2,0}(t)}{dt} &=&-\varepsilon _{2}P_{2,0}(t)+\varepsilon
_{2}P_{1,0}(t), \\
\frac{dP_{1,0}(t)}{dt} &=&-\left( \pi _{1}+\varepsilon _{1}\right)
P_{1,0}(t)+\pi _{1}P_{2,0}(t)+\varepsilon _{1}
\end{eqnarray}%
since $\pi _{2}=0.$
Solving above equations with the initial conditions $P_{1,0}(0)=P_{2,0}(0)=0,$ we find the cargo survival function $\Psi (t )=1-P_{1,0}(t):$
\begin{equation}
\Psi (t ) = p_{1} e^{-k_{1}t} + p_{2} e^{-k_{2}t},
\label{survival}
\end{equation}
where
\begin{equation}
p_{1} = \frac{\varepsilon _{1}}{k_2} \left( \frac{k_2 - \varepsilon _{1}}{k_{2}-k_{1}} \right),   \, \,   \,\,\,\,
p_{2} = \frac{\varepsilon _{1}}{k_1} \left( \frac{\varepsilon _{1} - k_{1}}{k_{2}-k_{1}} \right).
\end{equation}
Here two real eigenvalues $k_{1}$ and $k_{2}$ ($k_{1} < k_{2}$) are the solution of the quadratic equation
\begin{equation}
k^2 - (\pi _{1}+\varepsilon _{1} +\varepsilon _{2})k+ \varepsilon _{1}  \varepsilon _{2} =0.
\end{equation}
Since
\begin{equation}
p_{1} +
p_{2} = 1
\end{equation}
the survival function (\ref{survival}) has an interesting probabilistic interpretation 
if $\varepsilon _{2}-k_{1}>0$ and $k_{2}-\varepsilon _{2} >0$ 
(both $p_{1}$ and $p_{2}$ are positive). The cargo movement can be interpreted 
as one that involves a mixed population of two motors with different properties.
The first motor has a probability $p_{1}$, to be engaged and it has the exponential 
density of dwelling times with a rate $k_1$. $p_{2}$ is the probability of engagement 
for the second type of motor with an effective detachment rate $k_2.$
In this case the rate $\gamma (\tau )$ defined by (\ref{rate}) is {\it always} a decreasing function of running time $\tau$:
\begin{equation}
\gamma (\tau )=\frac{p_{1} k_{1} e^{-k_{1}\tau}+p_{2}k_{2} e^{-k_{2}\tau}}{p_{1} e^{-k_{1}\tau}+p_{2} e^{-k_{2}\tau}}.
\end{equation}
This explains the dramatic increase of run length of a cargo with two motors and non-Markovian nature of cargo movement. The detachment rate $\gamma (\tau )$ takes the maximum value at $\tau=0$:
\begin{equation}
\gamma (0)=p_{1} k_{1} +p_{2} k_{2} = \varepsilon _{1}.
\end{equation}
In the long time limit, the detachment rate $\gamma (\tau )$ tends to the constant value $k_{1}$ such that  $k_{1} < \varepsilon _{1}.$
This explains the dramatic increase of run length of a cargo with two motors
and the non-Markovian nature of cargo movement.
Our numerical results support this idea and show a decreasing detachment rate $\gamma$ (inset Fig.\ \ref{FIG2}).
Note that the empirical power-law survival function 
can be approximated with the survival function in the form of the linear combination of 
exponents corresponding to the Klumpp-Lipowsky model (Fig.\ \ref{FIG2}).

The survival function Eq.\ (\ref{survival}) has an interesting biological interpretation. 
The cargo movement can be viewed as one that involves a mixed population of two 
motors with different properties. If we extend this idea to a heterogeneous population 
of motors for which the rates $k$ are gamma distributed with the probability density 
function $f(k)=\tau_0 k^{\alpha-1} e^{-\tau_0 k}/\Gamma(\alpha)$, the effective 
survival function takes the form 
$\Psi(\tau)=\int_{0}^{\infty} e^{-k \tau} f(k) dk = (1+\tau/\tau_0)^{-\alpha}$ 
consistent with Eq.\ (\ref{Psi}).

\section{Non-Markovian dynamics of cargo: super-diffusion} \label{Sec4}

Our next aim is to provide a theoretical link
between the empirical mesoscopic detachment rate and the experimental L\'evy walk-like trajectories. 
The mesoscopic detachment rate Eq.\ (\ref{gamma}) with the anomalous exponent $1<\alpha<2$
applied in Eq.\ (\ref{str2}) allows us to explain the emergence of a L\'evy walk as a result of anomalous cumulative inertia
phenomena. We obtain the mean squared displacement (msd)
which exhibits sub-ballistic super-diffusive behaviour $\left<\mathbf{x}^{2}(t)\right> \sim t^{3-\alpha}$.
For L\'evy walk with $1<\alpha<2$ the ensemble and time averaged
msds differ only by a factor of $1/(\alpha-1)$ \cite{ZK,FB,MJCB}. 

Define the probability density function $\xi (t,\mathbf{x},\varphi ,\tau )$ which gives
the probability to find a cargo at point $\mathbf{x}=(x,y)$ at time $t$ that
moves with the velocity $v$ in the direction $\vect{\theta }=(\cos \varphi
,\sin \varphi )$ and having started the move a time $\tau$ ago.
Here $\varphi $ is the angle between the direction of movement $%
\vect{\theta }$ and the $x$ axis.
We assume that as long as the cargo detaches from the
microtubule it reattaches to another microtubule and thereby changes the
direction of movement.
The governing equation for $\xi (t,\mathbf{x},\varphi ,\tau )$ takes the
form \cite{Alt}
\begin{equation}
\frac{\partial \xi }{\partial t}+v\vect{\theta} \vect{\cdot \nabla }\xi +\frac{%
\partial \xi }{\partial \tau }=-\gamma (\tau )\xi,
\label{str2}
\end{equation}%
with the boundary condition
\begin{equation}
\xi (t,\mathbf{x},\varphi ,0)=\int_{0}^{t}\gamma (\tau )\int_{-\pi }^{\pi} 
R\left( \varphi - \varphi^{\prime }\right) \xi (t,\mathbf{x}, \varphi^{\prime },\tau) 
d\mathbf{\varphi }^{\prime } d\tau,
\label{in0}
\end{equation}%
where $R\left( \varphi - \varphi ^{\prime }\right) $ is the probability
density of the re-orientation from $\varphi ^{\prime }$ to $\varphi $ such
that $\int_{-\pi }^{\pi }R\left(u\right) du=1.$
In Ref.\ \cite{Granick} the flights were statistically isotropic with $R\left(u \right)=1/{2 \pi}$.
In our experiments, we observe quasi one-dimensional
trajectories (Fig.\ \ref{FIG1}) for which the angle takes only two values \cite{Fedotov16}.

The aim of this section is
to obtain the non-Markovian master equation for the probability density:
\begin{equation}
p(t,\mathbf{x},\varphi )=\int_{0}^{t}\xi (t,\mathbf{x},\varphi ,\tau )d\tau,
\label{den}
\end{equation}%
where the $\xi$ is the structural density.
We assume that at the initial time $t=0$ cargo has a zero running time
\begin{equation}
\xi (t,\mathbf{x},\varphi ,0)=p^{0}(t,\mathbf{x},\varphi )\delta (\tau ),
\label{initial}
\end{equation}
where $p^{0}(t,\mathbf{x},\varphi )$ is the initial density. 
The density $p(t,\mathbf{x},\varphi )$ can be found by differentiating (\ref{den}) with respect to
time $t$:
\begin{equation}
\frac{\partial p}{\partial t}+v\vect{\theta \cdot \nabla }p=-i(t,\mathbf{x}%
,\varphi )+j(t,\mathbf{x},\varphi ),  \label{eq1}
\end{equation}%
where $v$ is the cargo velocity, $\vect{\theta }=(\cos(\varphi), \sin(\varphi))$ is the direction of cargo's movement. 
The switching terms are:
\begin{equation}
i(t,\mathbf{x},\varphi )=\int_{0}^{t}\gamma (\tau )\xi (t,\mathbf{x},\varphi
,\tau )d\tau ,
\end{equation}%
\begin{equation}
 j(t,\mathbf{x},\varphi )=\xi (t,\mathbf{x},\varphi ,0)=
\label{sw}
\end{equation}%
$$
=\int_{0}^{t} \int_{-\pi }^{\pi }R\left(
\varphi - \varphi ^{\prime }\right) 
\gamma (\tau )
\xi (t,\mathbf{x},\varphi ^{\prime },\tau ) d\varphi ^{\prime } d\tau. 
$$
By using the method of characteristics we find  for $\tau <t$:
\begin{equation}
\xi (t,\mathbf{x},\varphi ,\tau )=\xi (t-\tau ,\mathbf{x-}v\vect{\theta }%
\tau ,\varphi ,0)e^{-\int_{0}^{\tau }\gamma (s)ds}.  \label{solution1}
\end{equation}%
The exponential factor in the above formula is the survival function:
\begin{equation}
\Psi (t)=e^{-\int_{0}^{t}\gamma (s)ds}.  \label{Sur}
\end{equation}%
To obtain $i(t,\mathbf{x},\varphi ),$ we use the
Fourier-Laplace transform:
\begin{equation}
\hat{\tilde{i}}(s,\mathbf{k},\varphi )=\int_{\mathbb{R}^{2}}\int_{0}^{\infty}i(t,%
\mathbf{x},\varphi )e^{i\vect{k\cdot x}-st} dt d\mathbf{x},  \label{LF1}
\end{equation}%
\begin{equation}
\hat{\tilde{p}}(s,\mathbf{k},\varphi )=\int_{\mathbb{R}^{2}}\int_{0}^{\infty}p(t,%
\mathbf{x},\varphi )e^{i \vect{k\cdot x}-st} dt d\mathbf{x}.  \label{LF2}
\end{equation}%
We find:
\begin{equation}
\hat{\tilde{i}}(s,\mathbf{k},\varphi )=\tilde{K}(s-iv\vect{k\cdot \theta })%
\hat{\tilde{p}}(s,\mathbf{k},\varphi ),  \label{LLL}
\end{equation}%
where $\tilde{K}(s)= \tilde {\psi} (s)/ \tilde {\Psi} (s).  $
Finally we obtain the expressions for the switching terms:
\begin{equation}
i(t,\mathbf{x},\varphi )=\int_{0}^{t}K(s)p(t-s,\vect{x}-v\vect{\theta }%
s,\varphi )ds,  \label{i}
\end{equation}%
\begin{equation}
j(t,\mathbf{x},\varphi )= \int_{-\pi }^{\pi }R\left(
\varphi - \varphi ^{\prime }\right) 
i(t,\mathbf{x},\varphi ^{\prime } )
d\varphi ^{\prime } .  \label{j}
\end{equation}%
The main advantage of the present derivation is that it can be easily
extended for the nonlinear case. 

Super-diffusive equations can be obtained for the case when the detachment 
rate $\gamma (\tau )$ can be approximated by the following rate
\begin{equation}
\gamma (\tau )=\frac{\alpha }{\tau _{0}+\tau },\ 1<\alpha <2.
\label{inverse}
\end{equation}%
The rate (\ref{inverse}) leads to a power law (Pareto) survival function:
\begin{equation}
\Psi (\tau )=\left[ \frac{\tau _{0}}{\tau _{0}+\tau }\right] ^{\alpha }
\label{Pareto}
\end{equation}%
and corresponding running time PDF:
\begin{equation}
\psi (\tau )=\frac{\alpha \tau _{0}^{\alpha }}{(\tau _{0}+\tau
)^{1+\alpha }}.  \label{eq:psi_tails}
\end{equation}%
The Laplace transform can be written in terms of the incomplete gamma
function $\Gamma (a,b)=\int_{b}^{\infty }t^{a-1}e^{-t}dt$ as:
\begin{equation}
\tilde{\psi}(s)=\alpha \left( \tau _{0}s\right) ^{\alpha }e^{\tau _{0}s}\Gamma
\left( -\alpha ,\tau _{0}s\right).   \label{La}
\end{equation}%
In the long-time limit as $s\rightarrow 0$, we have
\begin{equation}
\Gamma \left( -\alpha ,\tau _{0}s\right) =-\frac{\Gamma (1-\alpha )}{\alpha }%
+\left( \tau _{0}s\right) ^{-\alpha }\alpha ^{-1}+\frac{\left( \tau
_{0}s\right) ^{1-\alpha }}{1-\alpha }+...
\end{equation}

\begin{figure}[t]
\centerline{ \psfig{figure=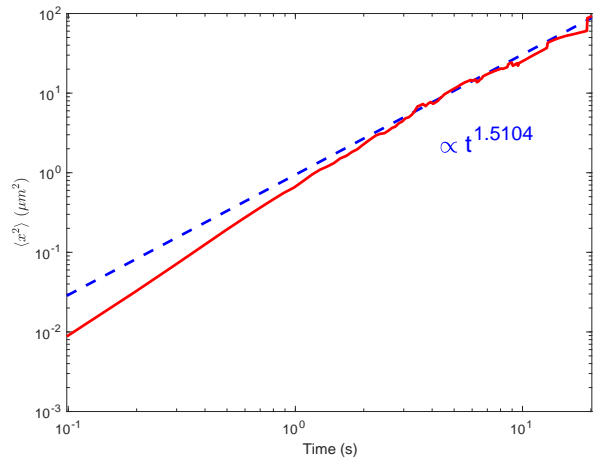,width=90mm,height=65mm} }
\caption{Anomalous behaviour of the time averaged mean squared displacement for 
experimental trajectories in RPE cell (corrected for the drift and averaged over many trajectories). 
The dashed curve represents the power law function with the exponent $1.5 \pm 0.17$.}
\label{FIG6}
\end{figure}

For $1<\alpha <2,$ using $\Gamma \left( -\alpha \right) =\frac{\Gamma
(2-\alpha )}{\alpha \left( \alpha -1\right) }$ we obtain:
\begin{equation}
\tilde{\psi}(s)\simeq 1-\frac{\tau _{0}s}{\alpha -1}+\frac{\Gamma (2-\alpha
)\left( \tau _{0}s\right) ^{\alpha }}{\left( \alpha -1\right) },\qquad
s\rightarrow 0
\end{equation}%
or
\begin{equation}
\tilde{\psi}\left( s\right) \simeq 1-{\overline T} s+\Gamma (2-\alpha )\tau
_{0}^{\alpha -1}{\overline T}s^{\alpha },  \label{small2}
\end{equation}%
where ${\overline T}=\tau _{0}\left( \alpha -1\right) ^{-1}$ is the mean value of
the random running time $T.$ Then
\begin{equation*}
\tilde{K}(s)=\frac{s\tilde{\psi}(s)}{1-\tilde{\psi}(s)}\simeq \frac{1}{{\overline T}}%
\left( 1+\Gamma (2-\alpha )\left( \tau _{0}s\right) ^{\alpha -1}\right)
\end{equation*}%
as $s\rightarrow 0.$ Using (\ref{LLL}) we write the Fourier-Laplace
transform of $i(t,\mathbf{x},\varphi )$ as:
\begin{equation}
\hat{\tilde{i}}=\frac{1}{{\overline T}}\left( 1+\Gamma
(2-\alpha )\tau _{0}^{\alpha -1}\left( s-iv\vect{k\cdot \theta }\right)
^{\alpha -1}\right) \hat{\tilde{p}}.  \label{sab}
\end{equation}%
The switching term $i(t,\mathbf{x},\varphi )$ can be written as \cite{SK}:
\begin{equation}
i=\frac{1}{{\overline T}}\left[ 1+\Gamma (2-\alpha )\tau
_{0}^{\alpha -1}\left( \frac{\partial }{\partial t}-v\vect{\theta \cdot
\nabla }\right) ^{\alpha -1}\right] p.
\label{i2}
\end{equation}%
where the fractional material derivative $\left(
\frac{\partial }{\partial t}-v\vect{\theta \cdot \nabla }\right)
^{\alpha -1 }$ of order $\alpha -1$ is defined by their Fourier-Laplace transforms
\begin{equation}
\mathcal{LF}\left\{ \left( \frac{\partial }{\partial t}-v\vect{\theta
\cdot \nabla }\right) ^{\alpha -1 }p\right\} =(s-iv\vect{k\cdot \theta }%
)^{\alpha -1}\tilde{p}(s,\mathbf{k},\varphi ).  \label{op}
\end{equation}%
Taking the Laplace transform of Eq.\ (\ref{eq1}), substituting the Laplace transform of Eq.\ (\ref{i2}) 
into it (using Eq. (\ref{j})) and Eq. (\ref{sab}), we solve the equation for $\tilde{p}$. 
Then, differentiating $\tilde{p}$ twice and taking the inverse Laplace transform, we 
find the mean squared displacement $\left<\mathbf{x}^{2}(t)\right>$.
For the isotropic case with the uniform angle distribution $R\left(
u \right)=1/{2 \pi}$ (Granick data \cite{Granick}) or quasi one-dimensional distribution (our data)
the mean squared displacement exhibits sub-ballistic
super-diffusive behaviour
\begin{equation}
\left<\mathbf{x}^{2}(t)\right>\sim t^{3-\alpha
},\qquad 1<\alpha <2.
\end{equation}
In Fig.\ \ref{FIG6} we show the time averaged mean squared displacement calculated for single experimental 
trajectories (corrected for the drift) which was also averaged over many trajectories. A clear power-law behaviour with 
the exponent $1.51 \pm0.17$ is found which is consistent with the L\'evy walk behaviour and 
with the behaviour of the detachment rate and survival functions.

\section{ Experimental Methods and Analysis of Trajectories} \label{Sec5}
 
Intracellular vesicles of U2OS (human bone osteosarcome epithelial) and RPE 
(retinal pigment epithelial) cells were imaged using phase contrast microscopy 
and tracked with Polyparticle Tracker software \cite{rogers2007}. 
The cells were grown in DMEM (Sigma Life Science) and 10\% FBS (HyClone) 
and incubated for 48 hrs at 37$^\circ$C in 8\% CO$_2$ on 35 mm glass-bottomed Ibidi dishes. 
Before imaging, the cells were moved to a live-imaging media. The live-cell imaging 
was done using an inverted Olympus IX71 with an Olympus 100$\times$/1.35 oil PH3 
objective and 1.6$\times$ zoom. A QuantEM 512SC CCD Camera and Cool LED pE-100 
light source was used for the continuous imaging of U2OS cells and a CoolSNAP 
HQ2 was used for the RPE cells. The video was taken with 30 ms or 98.5 ms exposure 
times while the cells were kept at 37$^\circ$ in atmospheric CO$_2$ levels.

After tracking each vesicle's path, only those with maximum displacements greater than 1 $\mu$m were chosen. 
Our aim was to filter those vesicles that are involved in active transport along microtubules. 
For this reason, we used the time averaged mean square displacements (MSDs) for single vesicle trajectories: 
\begin{equation}
\langle \mathbf{x}^2(m \delta t) \rangle = \frac{1}{N-m}\sum_{i = 1}^{N-m} (\mathbf{x}(t_i+m\delta t) - \mathbf{x}(t_i))^2,
\end{equation}
where $\mathbf{x}=(x,y)$ is the vesicle coordinate and the video contains $N$ 
snapshots at increments of $\delta t$. The total time of a data set is then 
$T = (N-1)\delta t$ and $m = 1,2,...,N-1$. Lag-times were defined as the set of 
possible $m\delta t$ within the data set.
Trajectories with mean square displacements (MSDs) close to $t^{2}$ (active transport) 
were analysed \cite{Arcizet}. The lower limit for the MSD was balanced between sufficient 
statistics to produce good fits and its proximity to $t^{2}$ proportionality. 

\begin{figure}[t]
\centerline{ \psfig{figure=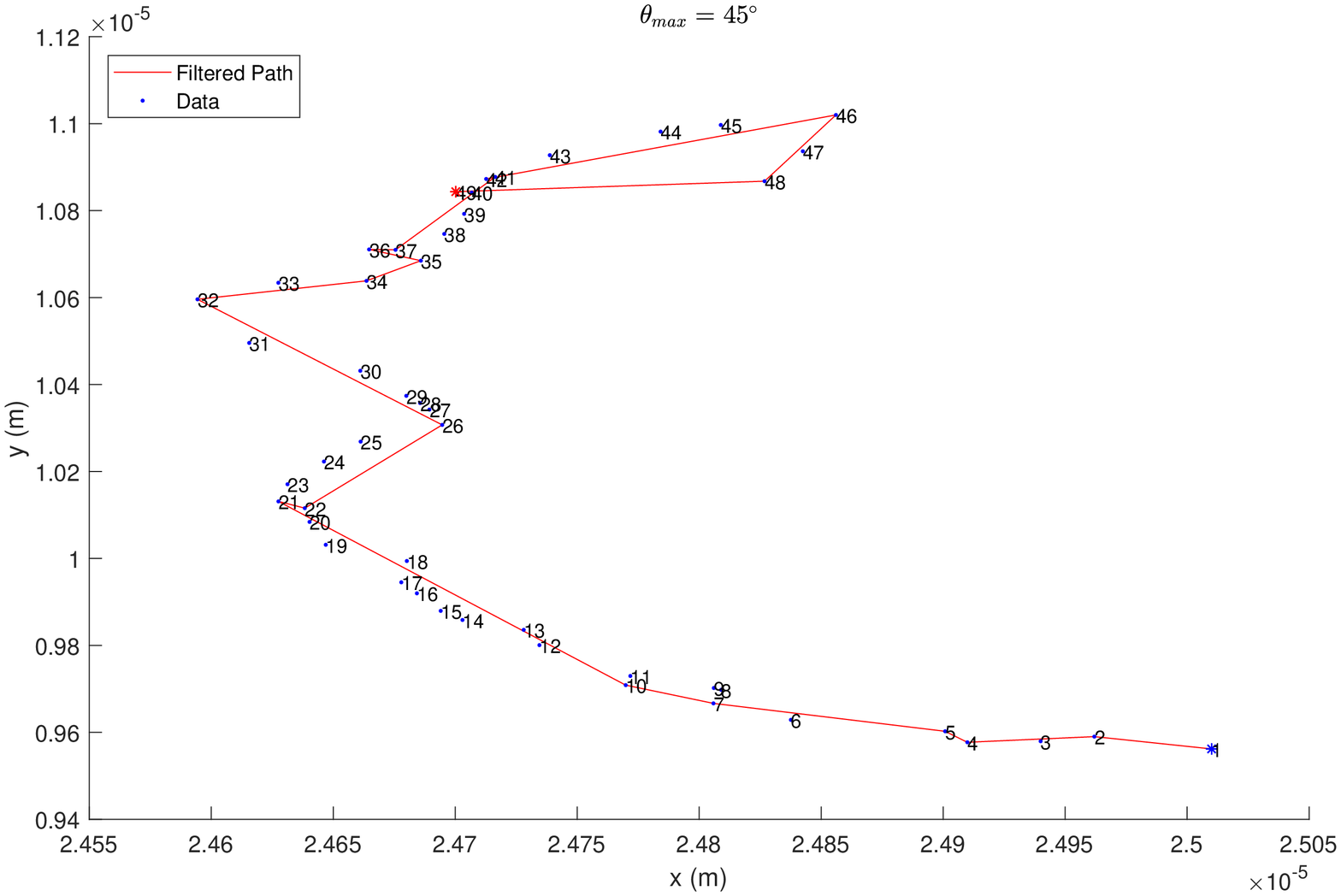,width=90mm,height=65mm} }
\caption{The Cartesian position of a tracked particle path segmented using $\theta_{\text{max}} = 45^{\circ}$. Blue
dots with frame numbers mark the path. The red line indicates the new segmented path. The blue
star marks the start position and the red star marks the end position.}
\label{FIGS1}
\end{figure}

To measure the turning times of the vesicles, an angular threshold method was used in 
conjunction with a distance threshold. In a set of $N$ points, there will be $N-2$ angles 
characterizing the direction of the path and for three arbitrary data points, $i$, $i+1$ and $i+2$, 
the angle deviation of the path was calculated by \cite{Harrison}:
\begin{equation}
\theta_{i+1} = \arccos\left(     \frac{\mathbf{x}_{i\rightarrow (i+1)} \cdot \mathbf{x}_{(i+1)\rightarrow (i+2)} }{|\mathbf{x}_{i \rightarrow (i+1)}||\mathbf{x}_{(i+1) \rightarrow (i+2)}| } \right).
\end{equation}
If $\theta_{i+1} > \theta_{max}$, the path was deemed to have changed direction. 
Additionally, if the tracked vesicle had not moved greater than 10\% of the length of 
a pixel, during the turn, the path was deemed to not have changed direction. 
This threshold was to ensure that stationary or completely detached vesicles, that move 
diffusively, do not add false runs to the statistics. In order to ensure results were not 
dependent on the arbitrary value of $\theta_{\text{max}}$, a set of threshold angles 
were tested ranging from 5$^{\circ}$ to 90$^{\circ}$ in 5$^{\circ}$ increments. 
We found $\theta_{\text{max}}=45^{\circ}$ to be optimal. With such an angular threshold, 
we have a good quality of the trajectory segmentation as shown in Fig.\ \ref{FIGS1}.  

\section{ Numerical modelling} \label{Sec6}

In experimental trajectories we found retrograde (towards the cell centre) and 
anterograde (away from the cell centre) movement which suggest that vesicles are driven by
both kinesins and dyneins. However, we observe no long pauses between 
retrograde and anterograde movement (Fig.\ $1$) 
and conclude that it is unlikely that kinesins and dyneins were 
engaged in a tug-of-war \cite{WelteGross98,MullerKlumppLipowsky}. 
Instead, changes of the direction could be triggered by some regulatory 
mechanism which remains poorly understood \cite{Hancock,Leidel}.

To model the multi-motor cargo transport we consider two groups of kinesin and dynein motors.  
Only kinesins or dyneins are engaged in cargo movement at any time. 
There are some indications (although still debated in the literature) that 
motors are loaded on the cargo in pairs \cite{Johnston}. We suppose that there is an equal number of kinesins and dyneins.
The  groups can change after the number of engaged motors reaches zero and the 
cargo detaches from the microtubules.
We assume that initially the cargo attaches to the microtubule with a single
kinesin or dynein motor and moves along it pulled by $N(t)$ motors of one polarity. 
The number of engaged motors $N(t)$ varies from $1$ to $\bar{N}$. 
The motors detach and reattach with rates $\varepsilon _{n}$ and $\pi _{n}$. 
It is assumed that all the motors equally share the external load. 

The number of kinesins and dyneins decreases and increases with the rates 
$\varepsilon _{n}$ and $\pi _{n}$ give by \cite{KunwarMogilner}:
\begin{equation}
\varepsilon _{n}= n \varepsilon e^{\frac{F_v}{n F_{d}}},  \; \; \;  \pi _{n}=\pi (\bar{N}-n), 
\end{equation}
where $F_v$ is the force acting on the cargo which depends on its velocity $v$, 
$F_d$ is the corresponding detachment force for kinesin or dynein, $\pi$ is the binding 
constant for a single kinesin or dynein motor and $\varepsilon$ is their zero-load unbinding rate. 
The values of $\pi$, $\varepsilon$ and $\bar{N}$ were adjusted to match the experimental data.

The cargo pulled by $n$ motors has the velocity 
\begin{equation}
v(F_v) = v \left( 1 - \frac{F_v}{n F_s} \right), 
\label{vF}
\end{equation}
where $F_s$ is the motor stall force and $v$ is the load-free velocity. The force $F_v$ which 
is acting on the cargo of radius $r$ due to viscous resistance also depend on the velocity 
of the cargo $F_v=6 \pi \eta r v$. Substituting this into Eq.\ (\ref{vF}) and solving 
for the velocity, the consistent expression for the velocity is \cite{KunwarMogilner}:
\begin{equation}
v(F_v) = \frac{v}{1+(6 \pi \eta r v)/(n F_s)}. 
\end{equation}
In our experiment, the typical radius of the lipid bound vesicles was $0.5$ $\mu$m. 
The cytoplasm is estimated to be $1000$ times more viscous than the buffer \cite{Luby-Phelps}, 
$\eta=0.89$ Pa s. Below we give the set of parameters used in simulations shown in Fig. $2$ in the main text. 
We have used parameters for kinesins and dyneins from Ref. \cite{Kunwar}. 
For kinesins: $v=1$ $\mu$m/s, $\pi=1$ s$^{-1}$, $\varepsilon=2$ s$^{-1}$, $F_d=4$ pN, $F_s=5$ pN  
For dyneins: $v=1$ $\mu$m/s, $\pi=1$ s$^{-1}$, $\varepsilon=3$ s$^{-1}$, $F_d=0.87$ pN, $F_s=1.25$ pN.

\section{Discussion and Conclusion}

In summary, we studied experimentally the non-Markovian anomalous multi-motor intracellular transport of cargoes inside cells.
To our knowledge, we for the first time directly measured the mesoscopic detachment rate of cargoes from microtubules and
demonstrated that the origin of the anomalous non-Markovian behavior is the cumulative inertia phenomenon. 
We provided evidence for this phenomenon in both bone and retina epithelial cells, but it is expected to occur in
all cell types that use dyneins and kinesins to transport cargoes along microtubules. 
We proposed a mesoscopic model which explains the emergence of memory and non-Markovian 
behaviour in the intracellular cargo transport on a mesoscopic scale. We also demonstrated how this non-Markovian 
behaviour emerged from Markovian memoryless dynamics of multiple motors on a microscopic scale. 
At the same time we note that the microscopic model is inconvenient since it has multiple 
parameters which require difficult fine tuning. 
And more importantly, almost all parameters in the microscopic model are unknown {\it in vivo}. 
There is no experimental technique to measure them. On the contrary, our 
mesoscopic model is computationally cheap and has only two parameters which we measure directly in the experiment.
We believe that our model provides a complementary description of the intracellular 
transport on a mesoscopic scale that is better able to model the
experimentally observed memory effects. 

The impact of our work on the field of intracellular transport will be three-fold: firstly, for the first time
we experimentally show the non-Markovian nature of intracellular transport and memory effects on a
mesoscopic scale. Secondly, our results settle the controversy in the field of intracellular transport (see
the work of Granick and co-workers in Nature Materials 2015) about memory-less Markovian
dynamics on a microscopic scale and non-Markovian behavior and memory effects on a mesoscopic
scale. Thirdly, with our work we are shifting the paradigm of how the single-particle tracking
experiments could be analysed by introducing several new mesoscopic statistical quantities, such as
the mesoscopic detachment rate, the survival function and the mean residual time to remain on the
microtubule. We believe that these quantities are better for a description of the long time properties of
intracellular transport than the traditionally used mean squared displacements along single
trajectories. What is important, is that we show that these quantities are experimentally measurable, 
can be predicted and can be modelled in a self-consistent manner (improving our confidence in the
robustness of our analysis). Improved non-Markovian modelling will lead to more accurate quantitative analysis
of the kinetics of a huge range of active transport in cellular physiology. Such transport 
impacts on a vast range of cellular processes and their diseased states e.g. motor neuron disease and cancer. 

\begin{acknowledgments}
SF, NK, TAW and VJA acknowledge financial support from the EPSRC Grant No. EP/J019526/1. D.H. acknowledges
funding from the Wellcome Trust, Grant No. 108867/Z/15/Z.
\end{acknowledgments}

\end{document}